\documentstyle[11pt,newpasp,twoside,epsf]{article}
\begin{document}
\title{Parallelization Strategies for ALI Radiative Transfer in Moving Media}
 \author{E. Baron}
\affil{Dept. of Physics and Astronomy, University of Oklahoma, Norman,
OK 73019-0261 USA}
\author{Peter H.~Hauschildt}
\affil{Dept. of Physics and Astronomy and Center for Simulational
Physics, University of Georgia, Athens, GA USA}
\author{David Lowenthal}
\affil{Dept. of Computer Science
Physics, University of Georgia, Athens, GA USA}

\begin{abstract}
We describe the method we have used to parallelize our spherically
symmetric special relativistic short characteristics general radiative
transfer code PHOENIX. We describe some possible parallelization
strategies and show why they would be inefficient. We discuss the
multiple parallelization strategy techniques that we have adopted. We
briefly discuss generalizing these strategies to full 3-D (spatial)
radiation transfer codes.
\end{abstract}

\section{Introduction}

In general parallelization is a subject to be avoided (as nicely
described by P.~H\"oflich, this volume), however
in order to take advantage of the enormous computing power and vast
memory sizes of modern parallel supercomputers, 
allowing both much faster model calculations as well as
more detailed models, 
we have implemented a parallel version of the general model atmosphere
code {\tt PHOENIX} (Hauschildt \& Baron 1999 and references
therein). Since the code 
uses a modular design, we have implemented different parallelization
strategies for different modules in order to maximize the total parallel
speed-up of the code. Our implementation allows us to change
the load distribution onto different processor elements (PEs) both via
input files and 
dynamically during a model run, which gives a high degree of flexibility
to optimize the performance on a number of different machines and for a
number of different model parameters.

Since we have both large CPU and memory requirements we have 
implemented the parallel version of the code on  using the
{\tt MPI}\ message passing library (MPI Forum 1995). 

We have chosen to work with the {\tt MPI}\ message passing interface,
since it is both portable (public domain implementations of {\tt MPI}
are readily available cf.~Gropp et~al. 1996), running on dedicated
parallel machines and heterogeneous workstation clusters and it is
available for both distributed and shared memory architectures.  For
our application, the distributed memory model is in fact easier to
code than a shared memory model, since then we do not have to worry
about locks and synchronization, etc.\ on {\em small} scales and we,
in addition, retain full control over interprocess communication.
This is especially clear once one realizes that it is fine to execute
the same code on many PEs as long as it is not too CPU intensive,
and avoids costly communication. We have added a few simple OpenMP
directives to our code, but do not discuss SMP parallelization further
here.

 An alternative to an implementation with {\tt MPI} is an
implementation using High Performance Fortran ({\tt HPF}) directives
(in fact, both can co-exist to improve performance). However, the
process of automatic parallelization guided by the {\tt HPF}
directives is presently not yet generating optimal results because the
compiler technology is still very new. 
{\tt HPF} is also
more suited for problems that are purely data-parallel (SIMD problems)
and would not benefit much from a MIMD approach. An optimal {\tt HPF}
implementation of {\tt PHOENIX} would also require a significant
number of code changes in order to explicitly instruct the compiler
not to generate too many communication requests, which would slow down
the code significantly. The {\tt MPI} implementation requires only the
addition of a few explicit communication requests, which can be done
with a small number of library calls.

\section{Basic numerical methods}

%
The co-moving frame radiative transfer equation for spherically
symmetric flows can be found in e.g., Mihalas \& Mihalas (1984).
%
The equation of radiative transfer is a integro-differential equation,
since the differential operator on the left hand side acts on the
specific intensity $I_\nu$, but the right hand side contains the 
emissivity $\eta_\nu$ which in turn contains $J_\nu$, the zeroth
angular moment of 
$I_\nu$:
\[ \eta_\nu = \kappa_\nu S_\nu + \sigma_\nu J_\nu, \]
 and 
\[ J_\nu = 1/2 \int_{-1}^{1} d\mu\,  I_\nu, \]
where $S_\nu$ is the source function, $\kappa_\nu$ is the absorption
opacity, and $\sigma_\nu$ is the scattering opacity.
With
the assumption of time-independence $\frac{\partial I_\nu}{\partial t} =
0$ and a monotonic velocity field the transfer equation  becomes a
boundary-value problem in 
the spatial coordinate and an initial value problem in the frequency
or wavelength coordinate. The equation can be written in operator form
as:
\begin{eqnarray}
J_\nu = \Lambda_\nu S_\nu,
\end{eqnarray}
where $\Lambda$ is the lambda-operator. 

\section{Parallel radiative transfer}

\subsection{Strategy and implementation}

We use a modified version of the method discussed in Hauschildt (1992)
for the numerical solution of the special relativistic radiative
transfer equation (RTE) at every wavelength point (see also
Hauschildt, St\"orzer, \& Baron 1994 and Hauschildt \& Baron 1999).
This iterative scheme is based on the operator splitting approach. The
RTE is written in its characteristic form and the formal solution
along the characteristics is done using a piecewise parabolic
integration (PPM, this is the ``short characteristic method'' Olson \&
Kunasz 1987).  We use the exact band-matrix subset of the discretized
$\Lambda$-operator as the `approximate $\Lambda$-operator' (ALO) in
the operator splitting iteration scheme (see Hauschildt
et~al. 1994). This gives very good convergence and high speed-up when
compared to diagonal ALO's.

 The serial radiative transfer code has been optimized for superscalar
and vector computers and is numerically very efficient.  It is
therefore crucial to optimize the ratio of communication to
computation in the parallel implementation of the radiative transfer
method. In terms of CPU time, the most costly parts of the radiative
transfer calculation are the setup of the PPM interpolation
coefficients and the formal solutions (which have to be performed in
every iteration). The construction of a tri-diagonal ALO requires
about the same CPU time as a single formal solution of the RTE and is
thus not a very important contributor to the total CPU time required
to solve the RTE at every given wavelength point.

In principle, the computation of the PPM coefficients does not require
any communication and thus could be distributed arbitrarily between
the PEs. However, the formal solution is recursive along each
characteristic. Within the formal solution, communication is only
required during the computation of the mean intensities, $J$, as they
involve integrals over the angle $\mu$ at every radial point. Thus, a
straightforward and efficient way to parallelize the radiative
transfer calculation is to distribute sets of characteristics onto
different PEs.  As Auer (this volume) has emphasized, the solution
of the radiative transfer equation is independent along a ray and in
the sense that we treat each characteristic separately for the
solution to the RTE, this parallelization step is one of domain
decomposition. Figure~1 illustrates the characteristics for a typical
supernova model calculation.

\begin{figure}
\plotfiddle{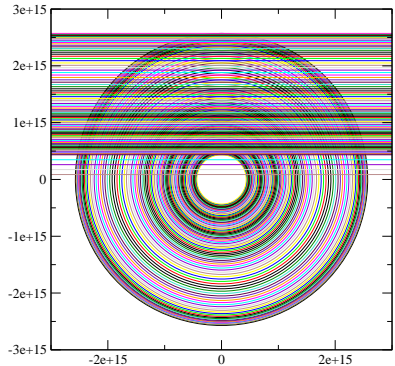}{0.6\hsize}{0}{30}{30}{-80}{0}
\caption{Schematic representation of the characteristic rays
in a typical supernova model calculation. The curvature of the rays
has been neglected in the figure, but is always included in the model
calculations.}
\end{figure}

Within one iteration
step, the current values of the mean intensities need to be broadcast
to all radiative transfer PEs and the new contributions of every
radiative transfer PE to the mean intensities at every radius must
be sent to the master PE. The master radiative transfer PE then
computes and broadcasts an updated $J$ vector using the operator
splitting scheme, the next iteration begins, and the process continues
until the solution is converged to the required accuracy. The setup,
i.e., the computation of the PPM interpolation coefficients and the
construction of the ALO, can be parallelized using the same method and
PE distribution. The communication overhead for the setup is roughly
equal to the communication required for a single iteration.

 An important point to consider is the load balancing between the
N$_{\rm RT}$ radiative transfer PEs. The workload to compute the
formal solution along each characteristic is proportional to the
number of intersection points of the characteristic with the
concentric spherical shells of the radial grid (the `number of points'
along each characteristic). Therefore, if the total number of points
is $N_{\rm P}$, the optimum solution would be to let each radiative
transfer PE work on $N_{\rm P}/N_{\rm RT}$ points. In general, this
optimum cannot 
be reached exactly  because it would require splitting
characteristics between PEs (which involves both communication and
synchronization). We therefore compromise by distributing the
characteristics to the radiative transfer PEs so that the total
number of points that are calculated by each PE is {\it roughly} the same
and that every PE works on a different set of characteristics.

\section{Parallelization of the line opacity calculation}

The contribution of spectral lines by both atoms and molecules is
calculated in {\tt PHOENIX}\ by direct summation of all contributing
lines.  Each line profile is computed individually at every radial
point for each line within a search window (typically, a few hundred
to few thousand Doppler widths or about $1000$\AA). This method is
more accurate than methods that rely on pre-computed line opacity
tables (the ``opacity sampling method'') or methods based on
distribution functions (the ``opacity distribution function (ODF)
method'').

 In typical model calculations we find that about 1,000 to 10,000
spectral lines contribute to the line opacity (e.g., in M dwarf model
calculations) at any wavelength point, requiring
a large number of individual Voigt profiles at every
wavelength point.  The subroutines for these computations can easily
be vectorized and we use a block algorithm with caches and direct
access scratch files for the line data to minimize storage
requirements. A block is the number of lines stored in active memory,
and the cache is the total number of blocks stored in memory. When the
memory size is exceeded, the blocks are written to direct access files
on disk. Thus the number of lines that can be included is not limited
by RAM, but rather by disk space and the cost of I/O. This approach is
computationally efficient because it provides high data and code
locality and thus minimizes cache/TLB misses and page faults. 

There are several obvious ways to parallelize the line opacity
calculations.  The first method is to let each PE compute the opacity
at $N_{\rm r}/N_{\rm PE}$ radial points, the second is to let each PE 
work on a different subset of spectral lines within each search window.
A third way, related to the second method, is to use completely
different sets of lines for each PE (i.e., use a global workload split
between the PEs in contrast to a local split in the second method).
We discuss the advantages and disadvantages of
these three methods. All three methods require only a very small 
amount of communication, namely a gather of all results to the
master PE with an {\tt MPI\_REDUCE} library call.

 The first and second methods, distributing sets of radial points or
sets of lines within the local (wavelength dependent) search window
over the PEs, respectively, are very simple to implement. They can
easily be combined to optimize their performance: if for any
wavelength point the number of depth points is larger than the number
of lines within the local search window, then it is more effective to
run this wavelength point parallel with respect to the radial points,
otherwise it is better to parallelize with respect to the lines within
the local window. It is trivial to add logic to decide the optimum
method for every wavelength point individually. This optimizes overall
performance with negligible overhead. The speed-up for this method of
parallelization is very close to the optimum value if the total number
of blocks of the blocking algorithm is relatively small ($\le 3\ldots
5$).

 However, if the total number of line blocks is larger (typically,
about 10 to 20 blocks are used), the overhead due to the read operations
for the block scratch files becomes noticeable and can reach
20\% or more of the total wall-clock time. This is due to the fact that
the `local parallelization' required that each PE working on the
line opacities needs to read every line block, thus increasing the
I/O time and load to the I/O subsystem by the number of PEs
themselves. We discuss I/O parallelization in \S~8.

\section{Parallelizing NLTE calculations}


Our method for iteratively solving the NLTE radiative transfer and
rate equations with an operator splitting method are discussed in
detail in Hauschildt (1993), and in Hauschildt \& Baron (1995, 1999),
therefore, we summarize our approach here only 
a briefly. The method uses a ``rate-operator'' formalism that
extends the approach of Rybicki \& Hummer (1991) to the general
case of multi-level NLTE calculations with overlapping lines and
continua, i.e, ``pre-conditioning''.  We use an ``approximate
rate-operator'' that is constructed 
using the exact elements of the discretized $\Lambda$-matrix (these
are constructed in the radiative transfer calculation for every
wavelength point). This approximate rate-operator can be either
diagonal or tri-diagonal. The method gives good convergence for a wide
range of applications and is very stable. It has the additional
advantage that it can handle very large model atoms, e.g., we use up
to 
11000 levels for NLTE model atoms in regular model calculations
(Hauschildt et~al. 1996, Baron et~al. 1996, Short et~al. 1999).

 Parallelizing the NLTE calculations involves parallelizing three
different sections: the NLTE opacities; the rates; and the solution of
the rate equations. In the following discussion, we consider only a
diagonal approximate rate-operator for simplicity. In this case, the
computation of the rates and the solution of the rate equations as
well as the NLTE opacity calculations can simply be distributed onto a
set of PEs without any communication (besides the gathering of the
data at the end of the iteration). This provides a very simple way of
achieving parallelism and minimizes the total communication
overhead. The generalization to a tridiagonal approximate rate-operator
is in principle straightforward, and involves only communication at the
boundaries between two adjacent PEs.

It would be possible to use the other two methods that we discussed
in the section on the LTE line opacities, namely local and global
set of lines distributed to different PEs. However, both would 
involve an enormous amount of communication between the NLTE PEs because
each NLTE transition can be coupled to any other NLTE transition. These
couplings require that a PE working on any NLTE transition have the 
required data to incorporate the coupling correctly. Although this
only applies to the PEs that work on the NLTE rates, it would require
both communication of each NLTE opacity task with each other
(to prepare the necessary data) and communication from the NLTE 
opacity PEs to the NLTE rate PEs. This could mean that several
MB data would have to be transferred between PEs at each
wavelength point, which is prohibitive with current
communication devices.

We have implemented the parallelization of the NLTE calculations by
distributing the set of radial points on different PEs. In order to
minimize communication, we also `pair' NLTE PEs so that each PE
works on NLTE opacities, rates, and rate equations for a given set of
radial points. This means that the communication at every wavelength
point involves only gathering the NLTE opacity data to the radiative
transfer PEs and the broadcast of the result of the radiative
transfer calculations (i.e., the ALO and the $J$'s) to the PEs
computing the NLTE rates. The overhead for these operations is
negligible but it involves synchronization (the rates can only be
computed after the results of the radiative transfer are known,
similarly, the radiative transfer PEs have to wait until the NLTE
opacities have been computed). Therefore, a good balance between the
radiative transfer tasks and the NLTE tasks is important to minimize
waiting. The rate equation calculations parallelize trivially over the
layers and involve no communication if the diagonal approximate
rate-operator is used. After the new solution has been computed, the
data must be gathered and broadcast to {\em all} PEs, the time for
this operation is negligible because it is required only once per
model iteration.

An important problem arises from the fact that the time spent in
the NLTE routines is not dominated by the floating point operations
(both the number and placement of floating point operation were
optimized in the serial version of the code) but by {\em memory access},
in particular for the NLTE rate construction. Although the parallel
version accesses a much smaller number of storage cells (which naturally
reduces the wall-clock time), effects like cache and TLB misses and page
faults contribute significantly to the total wall-clock time. All of
these can be reduced by using Fortran-90 specific constructs in the
following way: We have replaced the static allocation of the arrays that
store the profiles, rates etc.\ used in the NLTE calculations (done with
{\tt COMMON} blocks) with a Fortran-90 module and explicit allocation
(using {\tt ALLOCATE} and {\tt SAVE}) of these arrays at the start of
the model run. This allows us to tailor the size of the arrays to fit
exactly the number of radial points handled by each individual PE.
This reduces both the storage requirements of the code on every PE
and, more importantly, it minimizes cache/TLB misses and page faults. In
addition, it allows much larger calculations to be performed because the
RAM, virtual memory, and scratch disk space of every PE can be fully
utilized, thus effectively increasing the size of the possible
calculation by the number of PEs.

We find that the use of adapted array sizes significantly improves the
overall performance, even for a small number of processes. The scaling
with more PEs has also improved considerably. However, the overhead of
loops cannot be reduced and thus the speed-up cannot increase
significantly if more PEs are used. We have verified with a simple
test program which only included one of the loops important for the
radiative rate computation that this is indeed the case. Therefore, we
conclude that further improvements cannot be obtained at the Fortran-90
source level but would require either re-coding of routines in assembly
language (which is not practical) or improvements of the
compiler/linker/library system. We note that these performance changes
are very system dependent.

So far, (see also Hauschildt, Baron, \& Allard 1997) we have described our
method for parallelizing three separate modules: (1) The radiative
transfer calculation itself, where we divide up the characteristic
rays among processing elements and use an {\tt MPI\_REDUCE} to send
the $J_\nu$ to all the radiative transfer and NLTE rate computation
tasks;\footnote{We define a {\em task} as a logical unit of code that
treats an aspect of the physics of the simulation, such as radiative
transfer. On the other hand, we use the term {\em PE} to indicate a
single processing element of the parallel computer. Thus, a single
PE can execute a number of tasks, either serial on a single CPU or
in parallel, e.g., on an SMP PE of a distributed shared-memory
supercomputer.} (2) the line opacity which requires the calculation of
about 10,000 Voigt profiles per wavelength point at each radial grid
point, here we split the work amongst the processors both by radial
grid point and by dividing up the individual lines to be calculated
among the processors; and (3) the NLTE calculations.  The NLTE
calculations involve three separate parts: the calculation of the NLTE
opacities, the calculation of the rates at each wavelength point, and
the solution of the NLTE rate equations. So far our description has involved
parallelization by distribution of the radial grid points among
the different PEs or by distributing sets of spectral lines onto
different PEs. In addition, to prevent communication overhead, each
task computing the NLTE rates is paired on the same PE with and the
corresponding task computing NLTE opacities and emissivities to reduce
communication. The solution of the rate equations parallelizes
trivially with the use of a diagonal rate operator.

In the latest version of our code, {\tt PHOENIX 13.0}, we have incorporated
the additional strategy of distributing each NLTE species (the total
number of ionization stages of a particular element treated in NLTE) on
separate PEs. Since different species have different numbers of levels
treated in NLTE (e.g. Fe~II [singly ionized iron] has 617 NLTE levels,
whereas H~I has 30 levels), care is needed to balance the number of
levels and NLTE transitions treated among the PEs to avoid unnecessary
synchronization problems.

\section{Wavelength Parallelization}

The division of labor outlined in the previous sections requires
synchronization between the radiative transfer tasks and the NLTE
tasks, since the radiation field and ALO operator must be passed
between them. In addition, our standard model calculations use 50-100
radial grid points and as the number of PEs increases, so too does
the communication and loop overhead. 

Since the number of wavelength points in a calculation is very large
and the CPU time scales linearly with the number of wavelength points,
a further distribution of labor by wavelength points would potentially
lead to large speedups and to the ability to use very large numbers
of processors available on massively parallel supercomputers. Thus,
we have developed the concept of wavelength ``clusters'' to distribute
a set of wavelength points (for the solution of the frequency dependent
radiative transfer) onto a different set of PEs.  In order to achieve
optimal load balance and, more importantly, in order to minimize 
memory requirements, each cluster works on a single wavelength point
at a time, but it may consist of a number of ``worker'' PEs where the
worker PEs use parallelization methods discussed above.
In order to avoid communication overhead, the workers
on each wavelength cluster are {\em symmetric}: each corresponding worker
on each wavelength cluster performs identical tasks but on a different
set of wavelengths for each cluster. This allows us to make use of {\em
communicator contexts}, a concept which is built into {\tt MPI}. The
basic design is illustrated in Fig.~2. Each PE on a given
wavelength cluster is assigned to a {\tt MPI\_GROUP} and then the spatial
parallelization occurs within that {\tt MPI\_GROUP} so that, while the
communicators in different wavelength clusters have identical names,
they have different contexts and the code is much clearer and cleaner.
For example, Worker~0 will be working on the same set of characteristic
rays in the radiative transfer equation on all wavelength clusters.
Therefore, Worker~0 on Wavelength cluster~0 needs to send only the
intensities that it computed to its corresponding Worker~0 on Wavelength
cluster~1 and so forth. The code has been designed so that the number
of wavelength clusters, the number of workers per wavelength cluster,
and the task distribution within a wavelength cluster is arbitrary and
can be specified dynamically at run time.

\begin{figure}[ht]
\plotone{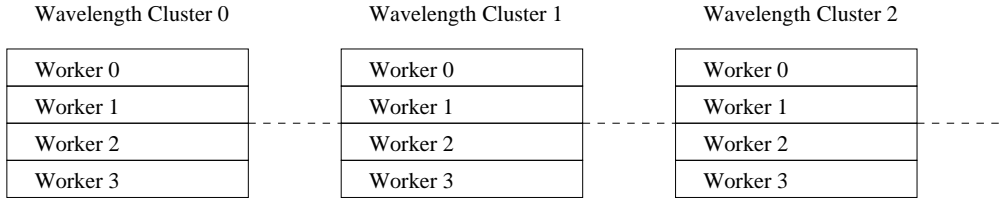}
\caption{The basic design of our parallelization
method, groups of processors are divided up into wavelength clusters
which will work on individual wavelength points, the wavelength
clusters are further divided into worker PEs, where each worker PE
is assign a set of specific (e.g., spatially distributed) tasks. Our design
requires that each worker PE on all wavelength clusters work on
exactly the same set of tasks, although additional inherently serial
operations can be assigned to one particular master worker, or master
wavelength cluster. This method reduces communication between clusters
to its absolute minimum and allows the maximum speedup.}
\end{figure}

In order to parallelize the spectrum calculations for a model atmosphere
with a global velocity field, such as the expanding atmospheres of
novae, supernovae or stellar  winds, we need to take the mathematical
character of the RTE into account. For monotonic velocity fields,
the RTE is an initial value problem in wavelength (with the initial
condition at the smallest wavelength for expanding atmospheres and
at the largest wavelength for contracting atmospheres). This initial
value problem must be discretized fully implicitly to ensure stability.
In the simplest case of a first order discretization, the solution 
of the RTE for wavelength point $i$ depends only on the results
of the point $i-1$. In order to parallelize the spectrum calculations,
the wavelength cluster computing the solution for wavelength point
$i$ must get the specific intensities from the cluster computing the
solution for point $i-1$. This suggests a ``pipeline'' solution to
the wavelength parallelization. Note that only the solution of
the RTE is affected by this, the calculation of the opacities and
rates remains independent between different wavelength clusters. In
this case, the parallelization works as follows: Each cluster can
independently compute the opacities, it then waits until it 
receives the specific intensities for the previous wavelength point,
computes the solution of the RTE and {\em immediately} sends the 
results to the next wavelength cluster (to minimize waiting time,
this is done with non-blocking send/receives), then proceeds to calculate
the rates and the new opacities for its {\em next} wavelength point
and so on. 

In the simplest case that there are simply 2 PEs and the load
balance has been set so that each PE acts as a wavelength cluster,
we have one worker per wavelength cluster and two wavelength clusters.
In this case,
PE 1 and 2 begin by doing the various preprocessing required.
PE 1 then begins working on wavelength point 1, PE 2 is working on
wavelength point 2. Since PE 1 begins with the boundary condition,
it calculates the opacities at this wavelength point and other
necessary preprocessing. It then solves the RTE, using the boundary
condition and then immediately sends the specific intensities off to
PE 2. PE 1 then calculates the rates and various other
post-processing at this wavelength point. At the same time PE 2 has
finished calculating the opacities at its wavelength point, done the
preprocessing for solving the RTE and then must wait until PE 1 has
sent it the specific intensities for the previous wavelength point at
which point it can then solve the RTE and immediately send its
specific intensities back to PE 1, which will need them for the next
wavelength point, PE 2 then calculates the rates at its wavelength
point and the process repeats. Since PE 1 was busy doing
post-processing, it may in fact have the specific intensities from PE
2, just in time and so minimal waiting may be required and the process
proceeds in a round robin fashion.
The basic pre- and post-processing required are illustrated in the
pseudo-code of Figure 3.

\begin{figure}[ht]
\begin{verbatim}

  for i := 1 to NUMWAVELENGTHS
     ....
     atomicLineOpacity(...)
     molecularLineOpacity(...)
     nlteOpacity(...)
     radiativeTransfer(...)
     nlteUpdateRates(...)
     ....
  end

\end{verbatim}
\caption[]{Pseudo-code for the global layout of
{\tt PHOENIX}. The processing that is required before and after the
radiative transfer is indicated.}
\end{figure}

This scheme has some predictable properties similar to the performance
results for classical serial vector machines. First, for a very small
number of clusters (e.g., two), the speedup will be small because the
clusters would spend a significant amount of time waiting for the results
from the previous cluster (the ``pipeline'' has no time to fill). Second,
the speedup will level off for a very large number of clusters when the
clusters have to wait because some of the clusters working on previous
wavelength points have not yet finished their RTE solution, thus limiting
the minimum theoretical time for the spectrum calculation to roughly the
time required to solve the RTE for all the wavelength points together
(the ``pipeline'' is completely filled).  This means that there is a
``sweet spot'' for which the speedup to number-of-wavelength-clusters
ratio is optimal. This ratio can be further optimized by using the
optimal number of worker PEs per cluster, thus obtaining an optimal
number of total PEs. The optimum will depend on the model atmosphere
parameters, the speed of each PE itself and the communication speed,
as well as the quality of the compilers and libraries.

The wavelength parallelization has the drawback that it does not
reduce the memory requirement per PE compared to runs with a single
wavelength cluster. Increasing the number of worker PEs per cluster
will decrease the memory requirements per PE drastically, however,
so that large runs can use both parallelization methods at the same
time to execute large simulations on PEs with limited memory. On a
shared-memory machine with distributed physical memory (such as the
Origin), this scheme can also be used to minimize memory access latency.

\section{Direct Opacity Sampling}

In our model atmosphere code we have implemented and used direct
opacity sampling (dOS) for more than a decade with very good
results. During that time, the size of the combined atomic and
molecular line databases that we used has increased from a few MB to
$>11\,$GB. Whereas the floating point and memory performance of
computers has increased dramatically in this time, I/O performance has
not kept up with this speed increase.  Presently, the wall-clock times
used by the line selection and opacity modules are dominated by I/O
time, not by floating point or overall CPU performance.  Therefore,
I/O performance is today more important that it was 10 years ago and
has to be considered a major issue. The availability of large scale
parallel supercomputers that have effectively replaced vector machines
in the last 5 years, has opened up a number of opportunities for
improvements of dOS algorithms. Parallel dOS algorithms with an
emphasis on the handling of large molecular line databases are thus an
important problem in computational stellar atmospheres. These
algorithms have to be portable and should perform well for different
types of parallel machines, from cheap PC clusters using Ethernet
links to high performance parallel supercomputers. This goal is
extremely hard to attain on all these different systems, and we
compare two different parallel dOS algorithms 
and describe their performance on two very different parallel
machines.

There are a number of methods in use to calculate line opacities. The
classical methods are statistical and construct tables that are
subsequently used in the calculation. The Opacity Distribution
Function (ODF) and its derivative the k-coefficient method have been
used successfully in a number of atmosphere and opacity table codes
(e.g., Kurucz 1992). This method works well for opacity table and
model construction but cannot be used to calculate detailed synthetic
spectra. A second approach is the opacity sampling (OS) method
(e.g., Peytremann 1974). This is a statistical approach in which the
line opacity is sampled on a fine grid of wavelength points using
detailed line profiles for each individual spectral line. In classical
OS implementation, tables of sampling opacities are constructed for
given wavelengths grids and for different elements.  

In direct opacity sampling (dOS) these problems are avoided by
replacing the tables with a direct calculation of the total line
opacity at each wavelength point for all layers in a model atmosphere
(Hauschildt \& Baron 1999). In the dOS method the relevant lines
(defined by a 
suitable criterion) are first selected from master spectral line
databases which include all available lines. The line selection
procedure will typically select more lines than can be stored in
memory and thus temporary line database files are created during the
line selection phase. The file size of the temporary database can
vary, in theory, from zero to the size of the original database or
larger, depending on the amount of data stored for the selected lines
and their number. For large molecular line databases this can easily
lead to temporary databases of several GB in size. This is in part due
the storage for the temporary line database: its data are stored for
quick retrieval rather than in the compressed space saving format of
the master line databases. The number and identity of lines that are
selected from the master databases depends on the physical conditions
for which the line opacities are required (temperatures, pressures,
abundances for a model atmosphere) and thus the line selection has to
be repeated if the physical conditions change significantly. 

The temporary line databases are used in the next phase to calculate
the actual line opacity for each wavelength point in a prescribed
(arbitrary) wavelength grid. This makes it possible to utilize
detailed line profiles for each considered spectral line on arbitrary
wavelength grids. For each wavelength grid point, all (selected) lines
within prescribed search windows (large enough to include all possibly
important lines but small enough to avoid unnecessary calculations)
are included in the line opacity calculations for this wavelength
point. 

\section{I/O Parallel Algorithms}


There are currently a large number of significantly different types of
parallel machines in use, ranging from clusters of workstations or PCs
to teraflop parallel supercomputers. These systems have very different
performance characteristics that need to be considered in parallel
algorithm design.  For the following discussion we assume this
abstract picture of a general parallel machine: The parallel system
consists of a number of processing elements (PEs), each of which is
capable of executing the required parallel codes or sections of
parallel code. Each PE has access to (local) memory and is able to
communicate data with other PEs through a communication device.  The
PEs have access to both local and global filesystems for data storage.
The local filesystem is private to each PE (and inaccessible to other
PEs), and the global filesystem can be accessed by any PE.  A PE can
be a completely independent computer such as a PC or workstation (with
single CPU, memory, and disk), or it can be a part of a shared memory
multi-processor system.   We assume
that the parallel machine has both global and local logical filesystem
storage available (possibly on the same physical device). The
communication device could be realized, for example, by standard
Ethernet, shared memory, or a special-purpose high speed communication
network.

In the following description of the 2 algorithms that we consider here
we will make use of the following features of the line databases:
\begin{itemize}
\item master line databases:
\begin{enumerate}
\item are globally accessible to all PEs 
\item are are sorted in wavelength
\item can be accessed randomly in blocks of prescribed size (number of lines)
\end{enumerate}
\item selected line (temporary) databases:
\begin{enumerate}
\item the wavelength grid is known during line selection (not
absolutely required but helpful) 
\item have to be sorted in wavelength
\item can be accessed randomly in blocks of prescribed fixed size
(number of lines) 
\item are stored either globally (one database for all PEs) or locally
(one for each PE) 
\item are larger than the physical memory of the PEs
\end{enumerate}
\end{itemize}

\subsection{Global Temporary Files (GTF)}


The first algorithm we describe relies on global temporary databases
for the selected lines. This is the algorithm that was implemented in
the versions of {\tt PHOENIX}\ discussed above. In the
general case of $N$ available PE's, the parallel line selection
algorithm uses one PE dedicated to I/O and $(N-1)$ line selection
PEs. The I/O PE receives data for the selected lines from the line
selection PEs, assembles them into properly sorted blocks of selected
lines, and writes them into the global temporary database for later
retrieval. The $(N-1)$ line selection PEs each read one block from a
set of $(N-1)$ adjacent blocks of line data from the master database,
select the relevant lines, and send the necessary data to the I/O
PE. Each line selection PE will select a different number of lines, so
the I/O PE has to perform administrative work to construct sorted
blocks of selected lines that it then writes into the global temporary
database. The block sizes for the line selection PEs and for the
global temporary database created by the I/O PE do not have to be equal, but
can be chosen for convenience.  The blocks of the master line database
are distributed to the $(N-1)$ line selection PEs in a round robin
fashion. Statistically this results in a balanced load between the
line selection PEs due to the physical properties of the line data.

After the line selection phase is completed, the global temporary line
database is used in the line opacity calculations. If each of the $N$
PEs is calculating line opacities (potentially for different sets of
wavelengths points or for different sets of physical conditions), they
all access the temporary database simultaneously, reading blocks of
line data as required. In most cases of practical interest, the same
block of line data will be accessed by several (all) PEs at roughly
the same time. This can be advantageous or problematic, depending on
the structure of the file system on which the database resides. The
PEs also cache files locally (both through the operating system and in
the code itself through internal buffers) to reduce explicit disk
I/O. During the line selection phase the temporary database
is a write-only file, whereas during the opacity calculations the
temporary database is strictly read-only.

The performance of the GTF algorithm depends strongly on the
performance of the global file system used to store the temporary
databases and on the characteristics of the individual PEs. This issue
is discussed in more detail below.

\subsection{Local Temporary Files (LTF)}



The second algorithm we consider uses file systems that are local to
each PE; such local file systems exist on many parallel machines,
including most clusters of workstations.  This algorithm tries to take
advantage of fast communication channels available on parallel
computers and utilizes local disk space space for temporary line list
files.  This local disk space is frequently large enough for the
temporary line database and may have high local I/O performance.  In
addition, I/O on the local disks of a PE does not require any inter-PE
communication, whereas globally accessible filesystems often use the
same communication channel that explicit inter-PE communication
uses. The latter can lead to network congestion if messages are
exchanged simultaneously with global I/O operations.


For the line selection, we could use the algorithm described above
with the difference that the I/O PE would create a single global (or local)
database of selected lines.  After the line selection is finished, the
temporary database could then simply be distributed to all PEs, and
stored on their local disk for subsequent use.  This is
likely to be slower in all cases than the GTF algorithm.

Instead, we use a ``ring'' algorithm that creates the local databases
directly.  In particular, each of the $N$ PEs selects lines for one
block from the master database (distributed in round robin fashion
between the PEs).  After the selection for this one block is complete,
each PE sends the necessary data to it next neighbor; PE $i$ sends its
results to PE $i+1$ and 
simultaneously, receives data from the previous PE in the ring (the
ring is periodic, i.e., PE $N$ sends data to PE $0$).  This
can be easily realized using the {\tt MPI\_SENDRECV}
call, which allows the simultaneous sending and receiving of data for
each member of the ring.  Each PE stores the data it receives into a
buffer and the process is repeated until the all selected lines from
the $N$ blocks are buffered in all $N$ PEs.  The PEs then transfer the
buffered line data into their local temporary databases. This cycle is
repeated until the line selection phase is complete. The line opacity
calculations will then proceed in the same way as outlined above,
however, the temporary line databases are now local for each PE.

This approach has the advantage that accessing the temporary databases
does not incur any (indirect) Network File System (NFS) communication
between the PEs as each of them has its own copy of the
database. However, during the line selection phase a much larger
amount of data has to be communicated over the network between the PEs
because now each of them has to ``know'' all selected lines, not only
the I/O PE used in the first algorithm.

The key insight here is that low-cost parallel computers constructed
out of commodity workstations typically have a very fast communication
network (100 Mbs to 1 Gbs) but relatively slow NFS performance.  This
means that trading off the extra communication for fewer NFS disk
accesses in the LTF algorithm is likely to give better performance.

\subsection{Results}

The performance of the GTF and LTF algorithms will depend strongly on
the type of parallel machine used.  A machine using NFS with fast
local disks and communication is likely to perform better with the LTF
algorithm. However, a system with  a fast (parallel) filesystem and fast
communication can  perform better with the GTF algorithm.  In
the following we will consider test cases run on very different
machines:
\begin{enumerate}
\item A cluster of Pentium Pro 200MHz PCs with 64MB RAM, SCSI disks,
100Mbs full-duplex  
Ethernet communication network, running Solaris 2.5.1.
\item An IBM SP system with 200MHz Power3 CPUs, 512MB RAM per CPU, 133
MB/s switched 
communication network, 16 PE IBM General Purpose File System (GPFS) parallel 
filesystem, running AIX 4.3.
\end{enumerate}


We have run 2 test cases to analyze the behavior of the 
different algorithms on different machines. The small test
case was designed to execute on the PPro system. It uses
a small line database (about 550MB) with about 35 million lines
of which about 7.5 million lines are selected. The second
test uses a database  which is about 16 times larger  ($\sim 9$GB)
and also selects 16 times more lines. This large test
could not be run on the PPro system due to file size 
limitations and limited available disk space. The line
opacity calculations were performed for about 21,000 
wavelength points that are representative of typical 
calculations. The tests were designed for maximum I/O 
usage and are thus extreme cases. In practical
applications the observed scaling is comparable to or
better than that found for these tests and appears to 
follow the results shown here rather well.

The results for the line selection procedure on the 
PPro/Solaris system are shown in Fig.~4.
It is apparent from the figure that the GTF approach 
delivers higher relative speedups that translate into
smaller execution times for more than 2 PEs. For serial
(1 PE) and 2 PE parallel runs the LTF line selection 
is substantially faster than the GTF algorithm. The reason
for this behavior can be explained by noting that the 
access of the global files is done through NFS mounts
that use the same network as the MPI messages. Therefore,
$n-1$ PEs request different data blocks from the NFS
server (no process was run on the NFS server itself) 
and send their results to the I/O PE, which 
writes it out to the NFS server. In the LTF algorithm,
each PE reads a different input block from the NFS
server and then sends its results (around the ring) to all other PEs. 
Upon receiving data from its left neighbor, a PE
writes it to local disk. This means that the amount
of data streaming over the network can be as much
as twice as high for the LTF compared to the 
GTF algorithm. This increases the execution time
for the LTF approach if the network utilization is close 
to the maximum bandwidth. In this argument we have ignored 
the time required to write the data to local disks, which 
would make the situation worse for the LTF approach.

The situation is very different for the calculation of the line
opacities (which uses only the local temporary database), as also shown in
Fig.~4. Now the LTF approach scales well 
(up to the maximum of 8 available machines) whereas the GTF algorithm
hardly scales to more than 2 PEs. The absolute execution times for the
LTF approach are up to a factor of 4 smaller (more typical are factors
around 2) than the corresponding times for the GTF algorithm (the GTF
run with 8 PEs required roughly as much execution time as the LTF run
with 1 PE!). The reason for this is clearly the speed advantage of 
local disk I/O compared to NFS based I/O in the GTF code.  If more
PEs are used in the GTF line opacity approach, the network becomes
saturated quickly and the PEs have to wait for their data (the NFS
server itself was not the bottleneck).  The LTF approach will be
limited by the fact that as the number of PEs gets larger, the
efficiency of disk caching is reduced and more physical I/O operations
are required. Eventually this will limit the scaling as the execution
time is limited by physical I/O to local disks.

\begin{figure}[hb]
\begin{tabular}{ll}
\begin{minipage}{2.0in}
\plotfiddle{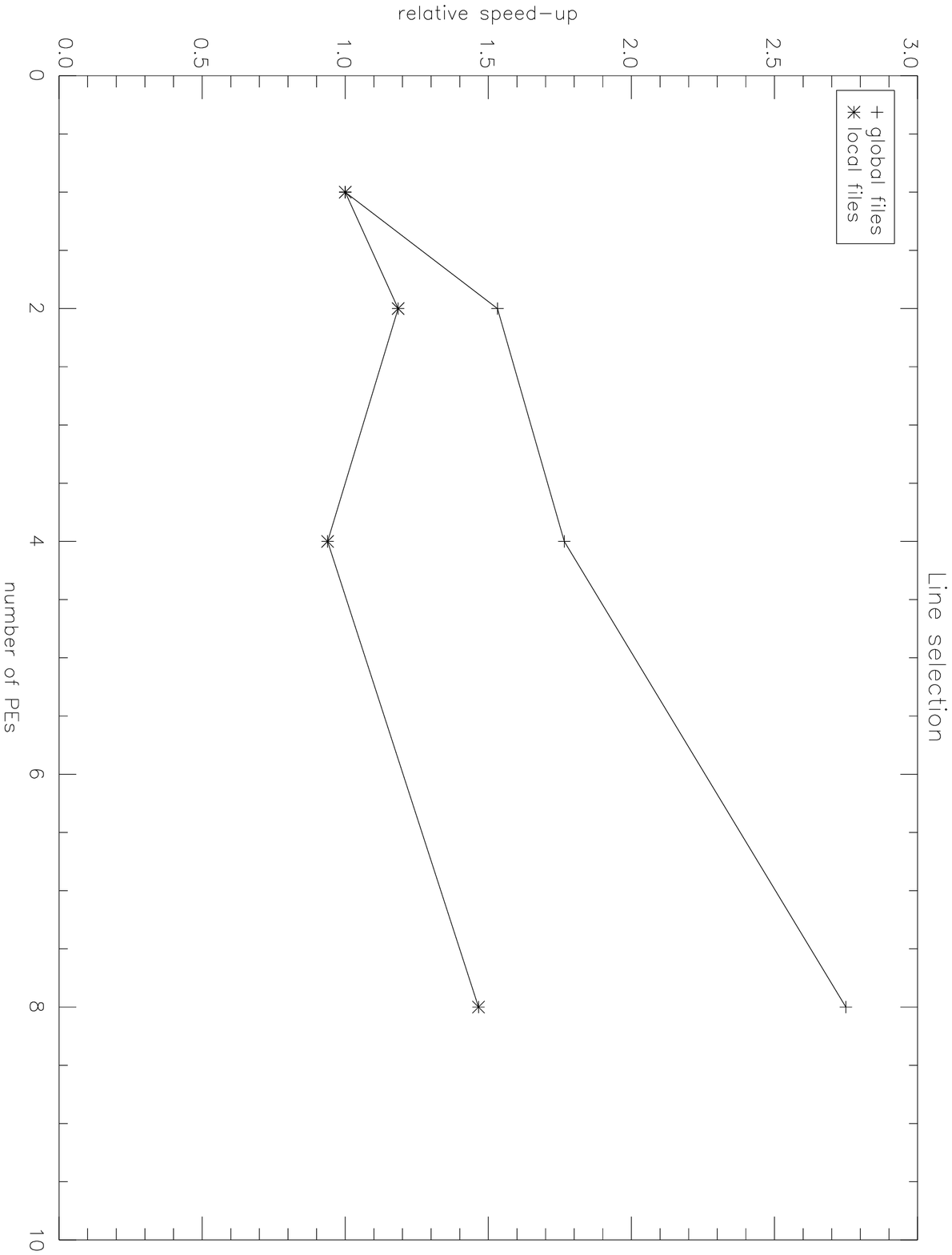}{2.0in}{90}{20}{20}{100}{0}
\end{minipage}
&
\begin{minipage}{2.0in}
\plotfiddle{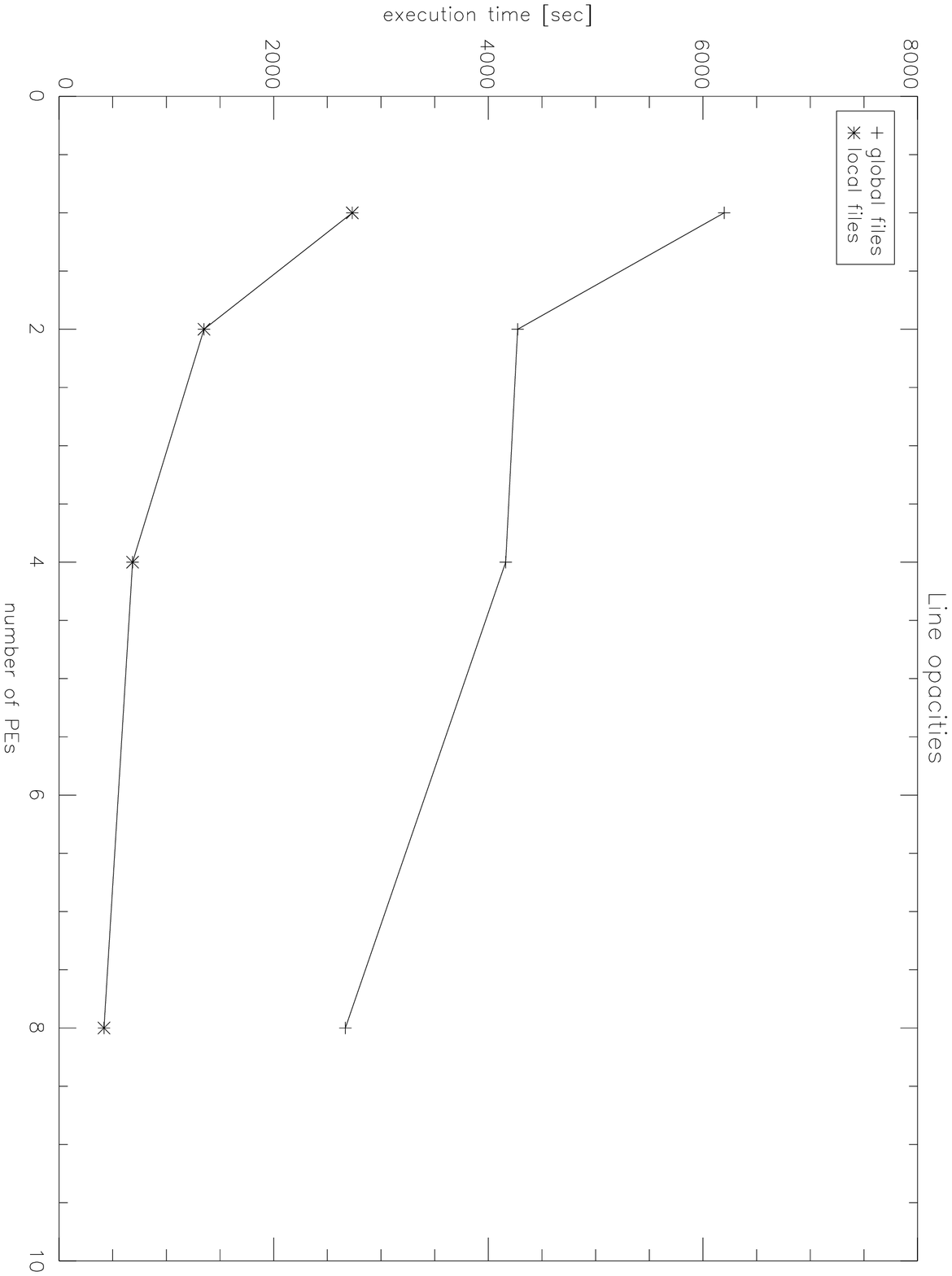}{2.0in}{90}{20}{20}{100}{0}
\end{minipage}
\end{tabular}
\caption[]{\label{ppro_line_sel}Relative speedups 
on the PPro/Solaris system. Results for the line selection are given
in the left panel, and for the line opacity calculation in the right
panel.}
\end{figure}


We ran the same (small) test on an IBM SP for comparison.  The tests
were run on a non-dedicated production system and thus timings are
representative of standard operation conditions and not optimum
values.  The global files were stored on IBM's General Purpose File
System (GPFS), which is installed on a number of system-dedicated I/O
PEs replacing the NFS fileserver used on the PPro/Solaris
system. GPFS access is facilitated through the same ``switch''
architecture that also carries MPI messages on the IBM SP.  
For the small test the results are markedly
different from the results for the PPro/Solaris system. The LTF
algorithm performs significantly better for all tested configurations,
however, scaling is very limited. The GTF code does not scale well at
all for this small test on the IBM SP.  This is due to the small size,
i.e., processing is so fast (nearly 100 times faster than on the
PPro/Solaris system) so that, e.g., latencies and actual line
selection calculations overwhelm the timing. The IBM SP has a very
fast switched communications network that can easily handle the higher
message traffic created by the LTF code. This explains why the LTF
line selection executes faster and scales better for this small test
on the IBM compared to the PPro/Solaris system.

The line opacity part of the test performs distinctively different in
the IBM SP compared to the PPro/Solaris system. 
In contrast to the latter, the IBM SP delivers
better performance for the GTF algorithm compared to the LTF code. The
scaling of the GTF code is also significantly better than that of the
LTF approach. This surprising result is a consequence of the high I/O
bandwidth of the GPFS running on many I/O PEs, the I/O bandwidth
available to GPFS is significantly higher than the bandwidth of the
local disks (including all filesystem overheads etc). The I/O PEs of
the GPFS can also cache blocks in their memory which can eliminate
physical I/O to a disk and replace access by data exchange over the
IBM ``switch''. Note that the test was designed and run with
parameters set to maximize actual I/O operations in order to
explicitly test this property of the algorithms.

The results of the large test case, for which the input file size is
about 16 times bigger, are very different for the line selection, as
shown in 
Fig.~5. Now the GTF algorithm executes much faster
(factor of 3) than the LTF code. This is probably caused by the larger
I/O performance of the GPFS that can easily deliver the data to all
PEs and the smaller number or messages that need to be exchanged in
the GTF algorithm. The drop of performance at 32 PE's in the GTF line
selection run could have been caused by a temporary overload of the
I/O subsystem (these tests were run on a non-dedicated machine).  In
contrast to the previous tests, the LTF approach does not scale in
this case.  This is likely caused by the large number of relatively
small messages that are exchanged by the PEs (the line list master
database is the same as for the GTF approach, so it is read through
GPFS as in the GTF case). This could be improved, e.g., by
choosing larger block sizes for data sent via MPI messages, however,
this will have the drawback of more memory usage and larger messages
are more likely to block than small messages that can be stored within
the communication hardware (or driver) itself.

The situation for the line opacities is also shown in Fig.~5.
The scaling is somewhat worse due to increased physical I/O (the temporary
files are about 16 times larger as in the small test case).  This is
more problematic for the LTF approach which scales very poorly for larger
numbers of PEs as the maximum local I/O bandwidth is reached far earlier than
for the GTF approach. This is rather surprising as conventional
wisdom would favor local disk I/O over global filesystem I/O on parallel 
machines. Although this is certainly true for farms of workstations or 
PCs, this is evidently not true on high-performance parallel computers 
with parallel I/O subsystems.

\begin{figure}[hb]
\begin{tabular}{ll}
\begin{minipage}{2.0in}
\plotfiddle{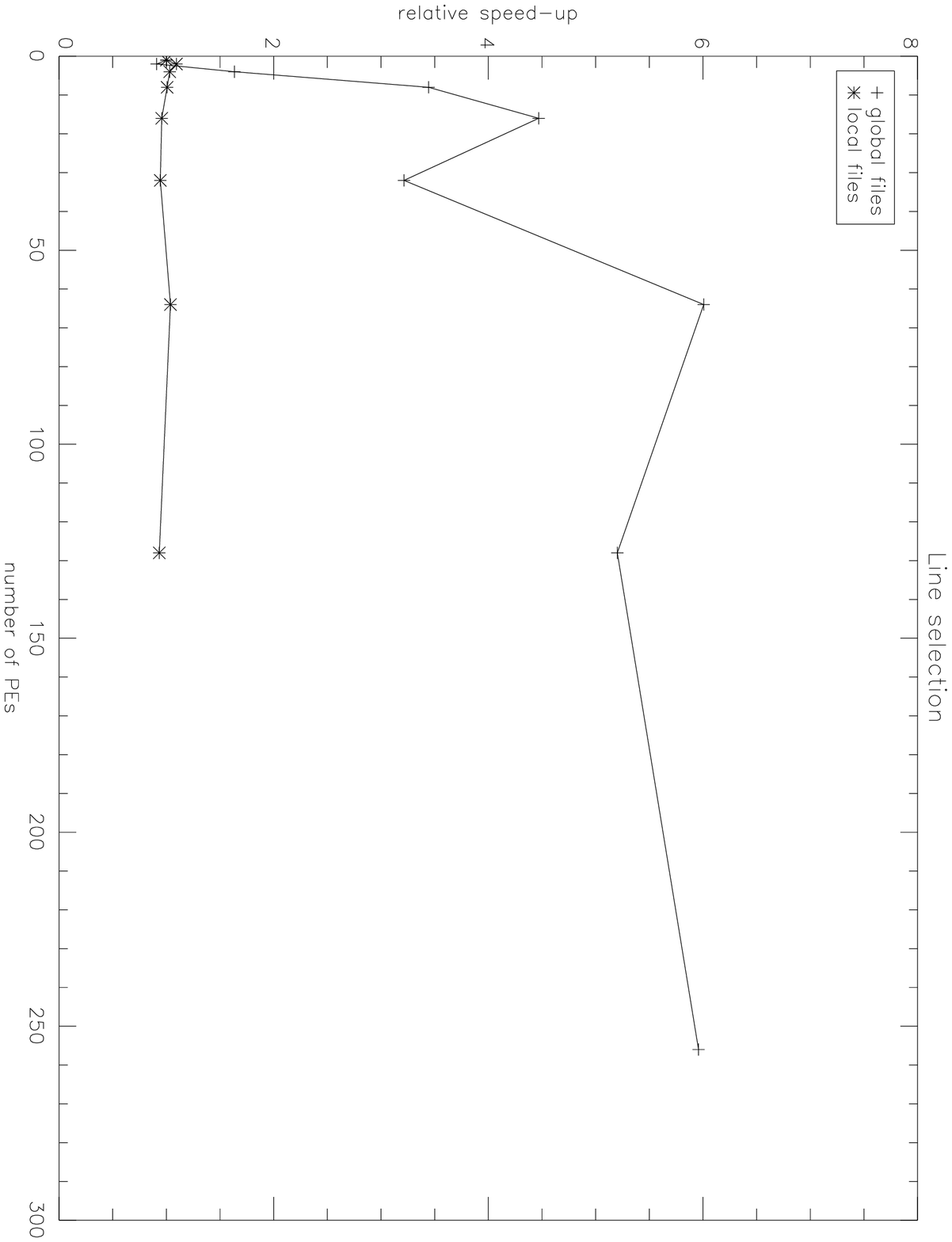}{0.6\hsize}{90}{20}{20}{100}{0}
\end{minipage}
&
\begin{minipage}{2.0in}
\plotfiddle{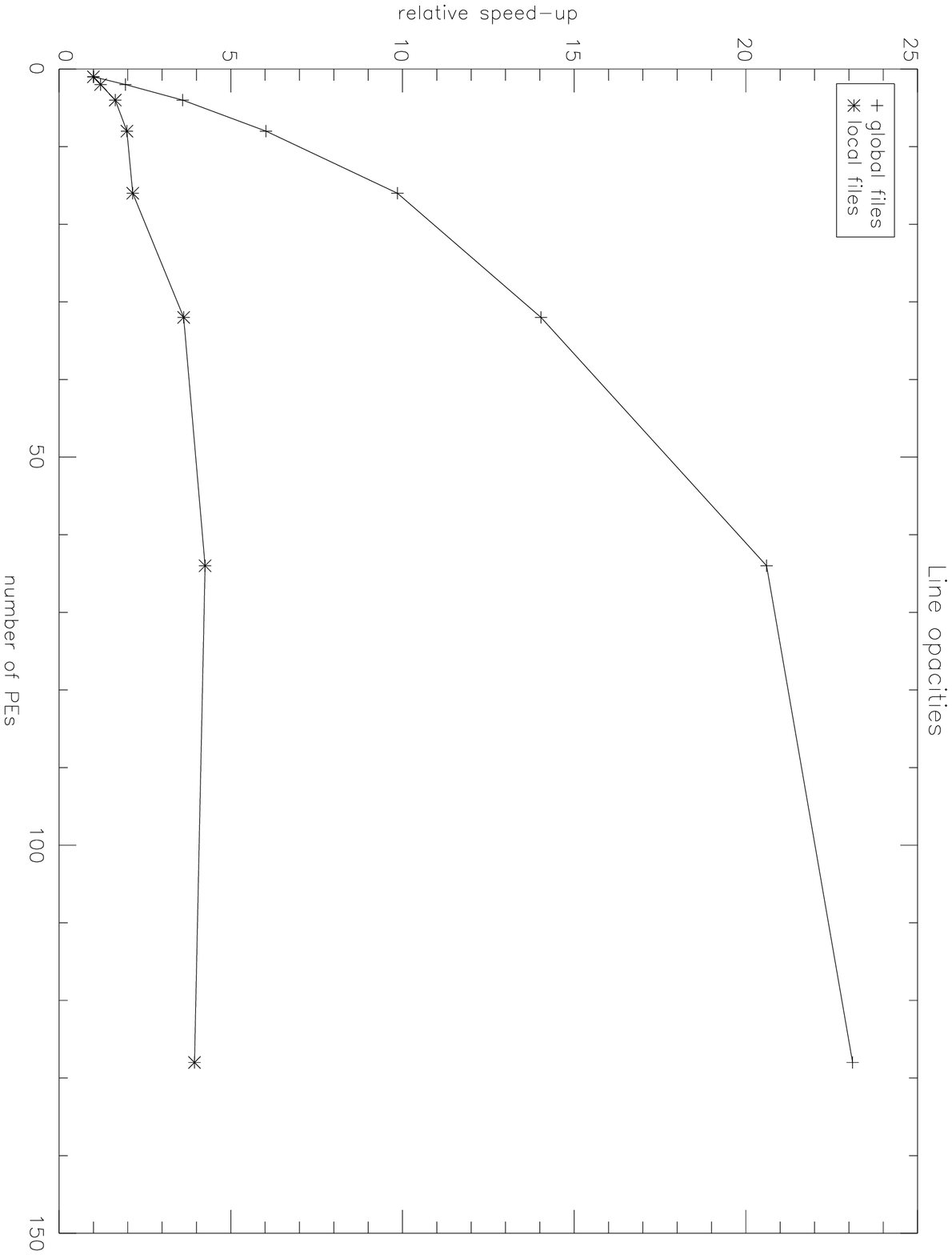}{0.6\hsize}{90}{20}{20}{100}{0}
\end{minipage}
\end{tabular}
\caption[]{\label{ibm_large_line_sel}Relative speedups for the 
large test 
on the IBM SP system.
Results for the line selection 
process are presented in the left panel and for the line opacity
calculations in
the right panel.}
\end{figure}


\clearpage

\section{Conclusions}

While the parallelization of a large numerical code such as {\tt
PHOENIX} would ideally be performed at the compiler level, it is not
possible with modern compilers to construct efficient code with good
scaling properties. Thus, individual modules of the code have to be
examined and parallelized separately. With {\tt MPI} this task is
manageable and allows the efficient use of both commodity parallel
clusters and modern massively parallel supercomputers.

We found good speedup up to about 64 PEs for a typical supernova
calculation.  However, for more than 64 PEs the communication,
synchronization, and loop overheads begin to become significant and it
is not economical to use more than 128 PEs 
For static
models and opacity tables we are able to use very large numbers of PEs
with scaling at close to the theoretical maximum. Fully 3-D
calculations will present yet another parallelization challenge.

\acknowledgments
 We thank our many collaborators who have
contributed to the development of {\tt PHOENIX}, in particular France Allard,
Jason Aufdenberg, Andreas Schweitzer, Travis Barman, Eric Lentz, Ian
 Short, David
Branch, Sumner Starrfield, and Steve Shore. 
This work was supported in part by NASA, the NSF, IBM, and the P\^ole
Scientifique de Mod\'elisation Num\'erique at ENS-Lyon and the
calculations were performed at NERSC and SDSC.

\end{document}